\documentclass[preprint]{aastex62}
\usepackage{booktabs}
\usepackage{graphicx}
\usepackage{blindtext}

\newcommand\ltsima{$\; \buildrel <\over\sim \;$}
\newcommand\simlt{\lower.5ex\hbox{\ltsima}}
\newcommand\gtsima{$\; \buildrel >\over\sim \;$}
\newcommand\simgt{\lower.5ex\hbox{\gtsima}}

\begin{document}
\title{OGLE-2017-BLG-1049: Another microlensing event with a giant planet}

\correspondingauthor{Sun-Ju Chung}
\email{sjchung@kasi.re.kr, sherlock@kasi.re.kr}

\author{Yun Hak Kim}
\affiliation{Korea Astronomy and Space Science Institute, 776 Daedeokdae-ro, Yuseong-Gu, Daejeon 34055, Korea} 
\affiliation{University of Science and Technology, Korea, (UST), 217 Gajeong-ro, Yuseong-gu, Daejeon 34113, Korea}

\author{Sun-Ju Chung}
\affiliation{Korea Astronomy and Space Science Institute, 776 Daedeokdae-ro, Yuseong-Gu, Daejeon 34055, Korea}
\affiliation{University of Science and Technology, Korea, (UST), 217 Gajeong-ro, Yuseong-gu, Daejeon 34113, Korea}

\author{A. Udalski  }
\affiliation{Warsaw University Observatory, AI.~Ujazdowskie~4, 00-478~Warszawa, Poland}

\author{Ian A. Bond  }
\affiliation{Institute of Natural and Mathematical Science, Massey University, Auckland 0745, New Zealand} 

\author{Youn Kil Jung  }
\affiliation{Korea Astronomy and Space Science Institute, 776 Daedeokdae-ro, Yuseong-Gu, Daejeon 34055, Korea} 

\collaboration{and}
\noaffiliation

\author{Andrew Gould  }
\affiliation{Korea Astronomy and Space Science Institute, 776 Daedeokdae-ro, Yuseong-Gu, Daejeon 34055, Korea}
\affiliation{Department of Astronomy, Ohio State University, 140 W. 18th Ave., Columbus, OH 43210, USA}
\affiliation{Max-Planck-Institute for Astronomy, K{\"o}nigstuhl 17, 69117 Heidelberg, Germany}

\author{Michael D. Albrow  }
\affiliation{Department of Physics and Astronomy, University of Canterbury, Private Bag 4800 Christchurch, New Zealand} 

\author{Cheongho Han  }
\affiliation{Department of Physics, Chungbuk National University, Cheongju 28644, Korea}

\author{Kyu-Ha Hwang  }
\affiliation{Korea Astronomy and Space Science Institute, 776 Daedeokdae-ro, Yuseong-Gu, Daejeon 34055, Korea}

\author{Yoon-Hyun Ryu  }
\affiliation{Korea Astronomy and Space Science Institute, 776 Daedeokdae-ro, Yuseong-Gu, Daejeon 34055, Korea}

\author{In-Gu Shin  }
\affiliation{Korea Astronomy and Space Science Institute, 776 Daedeokdae-ro, Yuseong-Gu, Daejeon 34055, Korea}

\author{Yossi Shvartzvald }
\affiliation{Department of Particle Physics and Astrophysics, Weizmann Institute of Science, Rehovot 76100, Israel}

\author{Jennifer C. Yee}
\affiliation{Center for Astrophysics $|$ Harvard \& Smithsonian, 60 Garden St., Cambridge, MA 02138, USA}

\author{Weicheng Zang  }
\affiliation{Department of Astronomy and Tsinghua Centre for Astrophysics, Tsinghua University, Beijing 100084, China}

\author{Sang-Mok Cha  }
\affiliation{Korea Astronomy and Space Science Institute, 776 Daedeokdae-ro, Yuseong-Gu, Daejeon 34055, Korea}
\affiliation{School of Space Research, Kyung Hee University, Giheung-gu, Yongin, Gyeonggi-do, 17104, Korea}

\author{Dong-Jin Kim  }
\affiliation{Korea Astronomy and Space Science Institute, 776 Daedeokdae-ro, Yuseong-Gu, Daejeon 34055, Korea}

\author{Hyoun-Woo Kim  }
\affiliation{Korea Astronomy and Space Science Institute, 776 Daedeokdae-ro, Yuseong-Gu, Daejeon 34055, Korea}
\affiliation{Department of Astronomy and Space Science, Chungbuk National University, Cheongju 28644, Republic of Korea}

\author{Seung-Lee Kim  }
\affiliation{Korea Astronomy and Space Science Institute, 776 Daedeokdae-ro, Yuseong-Gu, Daejeon 34055, Korea} 
\affiliation{University of Science and Technology, Korea, (UST), 217 Gajeong-ro, Yuseong-gu, Daejeon 34113, Korea}

\author{Chung-Uk Lee  }
\affiliation{Korea Astronomy and Space Science Institute, 776 Daedeokdae-ro, Yuseong-Gu, Daejeon 34055, Korea} 

\author{Dong-Joo Lee  }
\affiliation{Korea Astronomy and Space Science Institute, 776 Daedeokdae-ro, Yuseong-Gu, Daejeon 34055, Korea} 

\author{Yongseok Lee  }
\affiliation{Korea Astronomy and Space Science Institute, 776 Daedeokdae-ro, Yuseong-Gu, Daejeon 34055, Korea}
\affiliation{School of Space Research, Kyung Hee University, Giheung-gu, Yongin, Gyeonggi-do, 17104, Korea}

\author{Byeong-Gon Park  }
\affiliation{Korea Astronomy and Space Science Institute, 776 Daedeokdae-ro, Yuseong-Gu, Daejeon 34055, Korea} 
\affiliation{University of Science and Technology, Korea, (UST), 217 Gajeong-ro, Yuseong-gu, Daejeon 34113, Korea}

\author{Richard W. Pogge  }
\affiliation{Department of Astronomy, Ohio State University, 140 W. 18th Ave., Columbus, OH 43210, USA}

\collaboration{(KMTNet Collaboration),}
\noaffiliation

\author{Radek Poleski  }
\affiliation{Warsaw University Observatory, AI.~Ujazdowskie~4, 00-478~Warszawa, Poland}

\author{Przemek Mr{\'o}z  }
\affiliation{Division of physics, Mathematics, and Astronomy, California institute of Technology, Pasadena, CA 91125, USA}

\author{Jan Skowron  }
\affiliation{Warsaw University Observatory, AI.~Ujazdowskie~4, 00-478~Warszawa, Poland}

\author{Michal K. Szyma{\'n}ski  }
\affiliation{Warsaw University Observatory, AI.~Ujazdowskie~4, 00-478~Warszawa, Poland}

\author{Igor Soszy{\'n}ski  }
\affiliation{Warsaw University Observatory, AI.~Ujazdowskie~4, 00-478~Warszawa, Poland}

\author{Pawel Pietrukowicz  }
\affiliation{Warsaw University Observatory, AI.~Ujazdowskie~4, 00-478~Warszawa, Poland}

\author{Syzmon Koz{\l}owski  }
\affiliation{Warsaw University Observatory, AI.~Ujazdowskie~4, 00-478~Warszawa, Poland}

\author{Krzysztof Ulaczyk  }
\affiliation{Warsaw University Observatory, AI.~Ujazdowskie~4, 00-478~Warszawa, Poland}
\affiliation{Department of Physics, University of Warwick, Gibbet Hill Road, Coventry, CV4~7AL,~UK}

\author{Krzysztof A. Rybicki  }
\affiliation{Warsaw University Observatory, AI.~Ujazdowskie~4, 00-478~Warszawa, Poland}

\author{Patryk Iwanek  }
\affiliation{Warsaw University Observatory, AI.~Ujazdowskie~4, 00-478~Warszawa, Poland}

\collaboration{(OGLE Collaboration),}
\noaffiliation

\author{Fumio Abe  }
\affiliation{Institute for Space-Earth Environmental Research, Nagoya University, Nagoya 464-8601, Japan}

\author{Richard Barry  }
\affiliation{Code 667, NASA Goddard Space Flight Center, Greenbelt, MD 20771, USA}

\author{David P. Bennett  }
\affiliation{Code 667, NASA Goddard Space Flight Center, Greenbelt, MD 20771, USA}
\affiliation{Department of Astronomy, University of Maryland, College Park, MD 20742, USA}

\author{Aparna Bhattacharya }
\affiliation{Code 667, NASA Goddard Space Flight Center, Greenbelt, MD 20771, USA}
\affiliation{Department of Astronomy, University of Maryland, College Park, MD 20742, USA}

\author{Martin Donachie  }
\affiliation{Department of Physics, University of Auckland, Private Bag 92019, Auckland, New Zealand}

\author{Hirosane Fujii  }
\affiliation{Department of Earth and Space Science, Graduate School of Science, Osaka University, Toyonaka, Osaka 560-0043, Japan}

\author{Akihiko Fukui  }
\affiliation{Department of Earth and Planetary Science, Graduate School of Science, The University of Tokyo, 7-3-1 Hongo, Bunkyo-ku, Tokyo 113-0033, Japan}
\affiliation{Instituto de Astrof\'isica de Canarias, V\'ia L\'actea s/n, E-38205 La Laguna, Tenerife, Spain}

\author{Yoshitaka Itow  }
\affiliation{Institute for Space-Earth Environmental Research, Nagoya University, Nagoya 464-8601, Japan}

\author{Yuki Hirao  }
\affiliation{Department of Earth and Space Science, Graduate School of Science, Osaka University, Toyonaka, Osaka 560-0043, Japan}

\author{Rintaro Kirikawa  }
\affiliation{Department of Earth and Space Science, Graduate School of Science, Osaka University, Toyonaka, Osaka 560-0043, Japan}

\author{Iona Kondo  }
\affiliation{Department of Earth and Space Science, Graduate School of Science, Osaka University, Toyonaka, Osaka 560-0043, Japan}

\author{Naoki Koshimoto  }
\affiliation{Department of Astronomy, Graduate School of Science, The University of Tokyo, 7-3-1 Hongo, Bunkyo-ku, Tokyo 113-0033, Japan}
\affiliation{National Astronomical Observatory of Japan, 2-21-1 Osawa, Mitaka, Tokyo 181-8588, Japan}

\author{Yutaka Matsubara  }
\affiliation{Institute for Space-Earth Environmental Research, Nagoya University, Nagoya 464-8601, Japan}

\author{Yasushi Muraki  }
\affiliation{Institute for Space-Earth Environmental Research, Nagoya University, Nagoya 464-8601, Japan}

\author{Shota Miyazaki  }
\affiliation{Department of Earth and Space Science, Graduate School of Science, Osaka University, Toyonaka, Osaka 560-0043, Japan}

\author{Cl{\'e}ment Ranc  }
\affiliation{Code 667, NASA Goddard Space Flight Center, Greenbelt, MD 20771, USA}

\author{Nicholas J. Rattenbury  }
\affiliation{Department of Physics, University of Auckland, Private Bag 92019, Auckland, New Zealand}

\author{Yuki Satoh  }
\affiliation{Department of Earth and Space Science, Graduate School of Science, Osaka University, Toyonaka, Osaka 560-0043, Japan}

\author{Hikaru Shoji  }
\affiliation{Department of Earth and Space Science, Graduate School of Science, Osaka University, Toyonaka, Osaka 560-0043, Japan}

\author{Takahiro Sumi  }
\affiliation{Department of Earth and Space Science, Graduate School of Science, Osaka University, Toyonaka, Osaka 560-0043, Japan}

\author{Daisuke Suzuki  }
\affiliation{Department of Earth and Space Science, Graduate School of Science, Osaka University, Toyonaka, Osaka 560-0043, Japan}

\author{Paul J. Tristram  }
\affiliation{University of Canterbury Mt. John Observatory, P.O.Box 56, Lake Tekapo 8770, New Zealand}

\author{Yuzuru Tanaka  }
\affiliation{Department of Earth and Space Science, Graduate School of Science, Osaka University, Toyonaka, Osaka 560-0043, Japan}

\author{Tsubasa Yamawaki  }
\affiliation{Department of Earth and Space Science, Graduate School of Science, Osaka University, Toyonaka, Osaka 560-0043, Japan}

\author{Atsunori Yonehara  }
\affiliation{Department of Physics, Faculty of Science, Kyoto Sangyo University, Kyoto 603-8555, Japan}

\collaboration{(MOA Collaboration)}
\noaffiliation

\begin{abstract}
We report a giant exoplanet discovery in the microlensing event OGLE-2017-BLG-1049, which is a planet-host star mass ratio of $q=9.53\pm0.39\times10^{-3}$ and has a caustic crossing feature in the Korea Microlensing Telescope Network (KMTNet) observations. The caustic crossing feature yields an angular Einstein radius of $\theta_{\rm E}=0.52 \pm 0.11\ {\rm mas}$. However, the microlens parallax is not measured because of the time scale of the event $t_{\rm E}\simeq 29\ {\rm days}$, which is not long enough in this case to determine the microlens parallax. Thus, we perform a Bayesian analysis to estimate physical quantities of the lens system. From this, we find that the lens system has a star with mass $M_{\rm h}=0.55^{+0.36}_{-0.29} \ M_{\odot}$ hosting a giant planet with $M_{\rm p}=5.53^{+3.62}_{-2.87} \ M_{\rm Jup}$, at a distance of $D_{\rm L}=5.67^{+1.11}_{-1.52}\ {\rm kpc}$. The projected star-planet separation in units of the Einstein radius $(\theta_{\rm E})$ corresponding to the total mass of the lens system is $a_{\perp}=3.92^{+1.10}_{-1.32}\ \rm{au}$. This means that the planet is located beyond the snow line of the host. The relative lens-source proper motion is $\mu_{\rm rel}\sim 7 \ \rm{mas \ yr^{-1}}$, thus the lens and source will be separated from each other within 10 years. Then the flux of the host star can be measured by a 30m class telescope with high-resolution imaging in the future, and thus its mass can be determined.
\end{abstract}

\section{Introduction}
Up to now, 4301 exoplanets (as of 5 November \ 2020)\footnote{https://exoplanetarchive.ipac.caltech.edu/index.html\label{nasa}} have been detected by various methods, such as radial velocity, transit, direct imaging, and microlensing. Over $90 \%$ of them have been discovered by the radial velocity and transit methods, and they are almost all located close to their host stars. Thus, the great majority of exoplanets are biased to close-in planets, i.e., hot planets. By contrast, almost all of the exoplanets discovered by microlensing are cold planets, which lie at or beyond the snow line i.e., where the water can form into ice in the proto-planetary disk (Kennedy \& Kenyon\citealt{kennedy2008}).

The majority of host stars of microlensing exoplanets are faint low-mass M dwarf stars, which are generally difficult to detect from radial-velocity and transit observations. This is because microlensing does not depend on the brightness of objects, only the mass (Gaudi\citealt{gaudi2012}). The microlensing host stars have many giant planets beyond the snow line, suggesting that giant planets around M dwarfs might be common (e.g., Gould et al.\citealt{gould2006}; Sumi et al.\citealt{sumi2010}; Montet et al.\citealt{montet2014}). This is contrary to the prediction of the core-accretion model of the planet formation for which the planet frequency around M dwarf should be rare (Ida \& Lin\citealt{ida2004}; Laughlin et al.\citealt{laughlin2004}; Kennedy et al.\citealt{kennedy2006}), while it is consistent with the prediction of the disk instability model for which they would be common (Boss\citealt{boss2006}). These results imply that microlensing exoplanets are very important to better understand planet formation. 

As of now, the total number of exoplanets detected by microlensing has reached 105 (as of 5 November, 2020)\textsuperscript{\ref{nasa}}, which is relatively small compared to the numbers detected by the radial velocity and transit methods. Of these, $42\%$ of the host stars have masses that are only estimated by a Bayesian analysis, which investigates the probability distributions of physical parameters (Jung et al.\citealt{jung2018}). Since M dwarfs are the most common stars in the Galaxy, this necessarily results in a prediction that most of the host stars are M dwarfs. At the same time, the Bayesian prior explicitly assumes that planet frequency is independent of host star mass.

In order to truly test planet formation theory, we need to assemble a sample of microlensing giant planets whose host stars have direct mass measurements. Often, such measurements can be made by resolving the light from the lens star using high-resolution imaging. For example, Dong et al.\citet{dong2009} showed that the giant planet in OGLE-2005-BLG-071 does indeed orbit an M-dwarf, a result that was later confirmed by Bennett et al.\citet{bennett2020}. By contrast, Vandorou et al.\citet{vandorou2020} measured the lens flux for MOA-2013-BLG-220, which shows a clear difference from the central value of the Bayesian analysis. Robust flux measurements are best made after the source and the lens have separated (Battacharya et al.\citealt{battacharya2017}), i.e. many years after the time of the event. Thus, the first step to a statistical study of giant planets around M dwarfs is to identify a large sample of giant planets for future high-resolution follow-up observations.

Until now, $54$ planetary systems composed of giant planets around M dwarf have been detected by microlensing. The microlensing event OGLE-2017-BLG-1049 is such the planetary system. In this paper, we report the analysis of the planetary event OGLE-2017-BLG-1049. The light curve of the event has a U-shaped caustic crossing feature. This feature occurs when the source crosses the caustic, which represents the set of source positions at which a magnification of a point source becomes infinite (Chung et al.\citealt{chung2005},\ Han\citealt{han2005}). The caustic crossing feature was covered by the Korea Microlensing Telescope Network (KMTNet; Kim et al.\citealt{kim2016}).  

The paper is organized as follows. Observations of the event are described in Section 2. In Section 3, the light curve analysis is described, and the physical parameters of the lens are estimated in Section 4. Then, we summarize the result in Section 5.

\section{Observation}
The microlensing event OGLE-2017-BLG-1049 was discovered on $2017$ June $4$ by the Optical Gravitational Lensing Experiment (OGLE; Udalski \citealt{udalski2003}) survey with the 1.3 m Warsaw telescope at Las Campanas Observatory in Chile. The equatorial and galactic coordinates of the event are $(\alpha, \delta)=(17^{\rm h}58^{\rm m}8^{\rm s} 05, -27^{\circ} 08^{'} 39^{''}20)$ and $(l, b)=(2^{\circ} 950,-1^{\circ} 461)$, respectively.

The event was also found by the KMTNet post-season event finder (Kim et al.\citealt{kim2018}) and was designated as KMT-2017-BLG-0370. KMTNet observations were conducted using three identical telescopes at the Cerro Tololo Inter-American Observatory in Chile (KMTC), South African Astronomical Observatory in South Africa (KMTS), and Siding Spring Observatory in Australia (KMTA). The event lies in two overlapping fields, BLG03 and BLG43, with a combined cadence of $\Gamma =4$ hr$^{-1}$. The majority of images were taken in $I$ band, and some data were taken in $V$ band to measure the color of the source star. The KMTNet data were reduced by pySIS pipeline based on the Difference Image Analysis (DIA; Alard $\&$ Lupton\citealt{alard1998}, Albrow et al.\citealt{albrow2009}). 
The event was originally analysed using OGLE and KMT data because it had not been alerted by the Microlensing Observations in Astrophysics (MOA; Sumi et al.\citealt{sumi2016}). The best-fit model without MOA predicts the second strong anomaly feature right in a gap in the data, which might be regarded as \textquotedblleft suspicious\textquotedblright (see Figure ~\ref{fig:fig1}). Review of the MOA observations showed that they had observed this field during the gap, so special reductions were carried out. Thus, the MOA data were included to confirm the second anomaly feature in the gap (see Figure ~\ref{fig:fig2}).

In general, each observatory has their own photometry pipeline packages and they typically underestimate the true error due to their systematics (Yee et al.\citealt{yee2012}).  Hence, the errors of each data in magnitude are required to normalize when they are used together in the modeling. Thus, we renormalize the error bars on each data sets except OGLE. For OGLE data sets, Skowron et al.\citet{skowron2016} already have done the renormalization for all their survey fields, and thus they provide the error correction coefficients. From this, we adopt the rescaling parameters for OGLE. For the renormalization, we follow the procedure (see Eq.~\ref{eqn:yee}) described in Yee et al.\citet{yee2012},
\begin{equation}
	\sigma'=k\sqrt{\sigma_{\it i}^2+(\sigma_{0})^2},
	\label{eqn:yee}
\end{equation}
where $\sigma_{\it i}$ is the original error bar of the {\it i}th data, and {\it k} and $\sigma_{0}$ are the rescaling parameters for making $\chi^{2}/$dof$\ =1$. The rescaling parameters are presented in Table 1.

\section{Light curve Analysis}
\begin{figure}
	\plotone{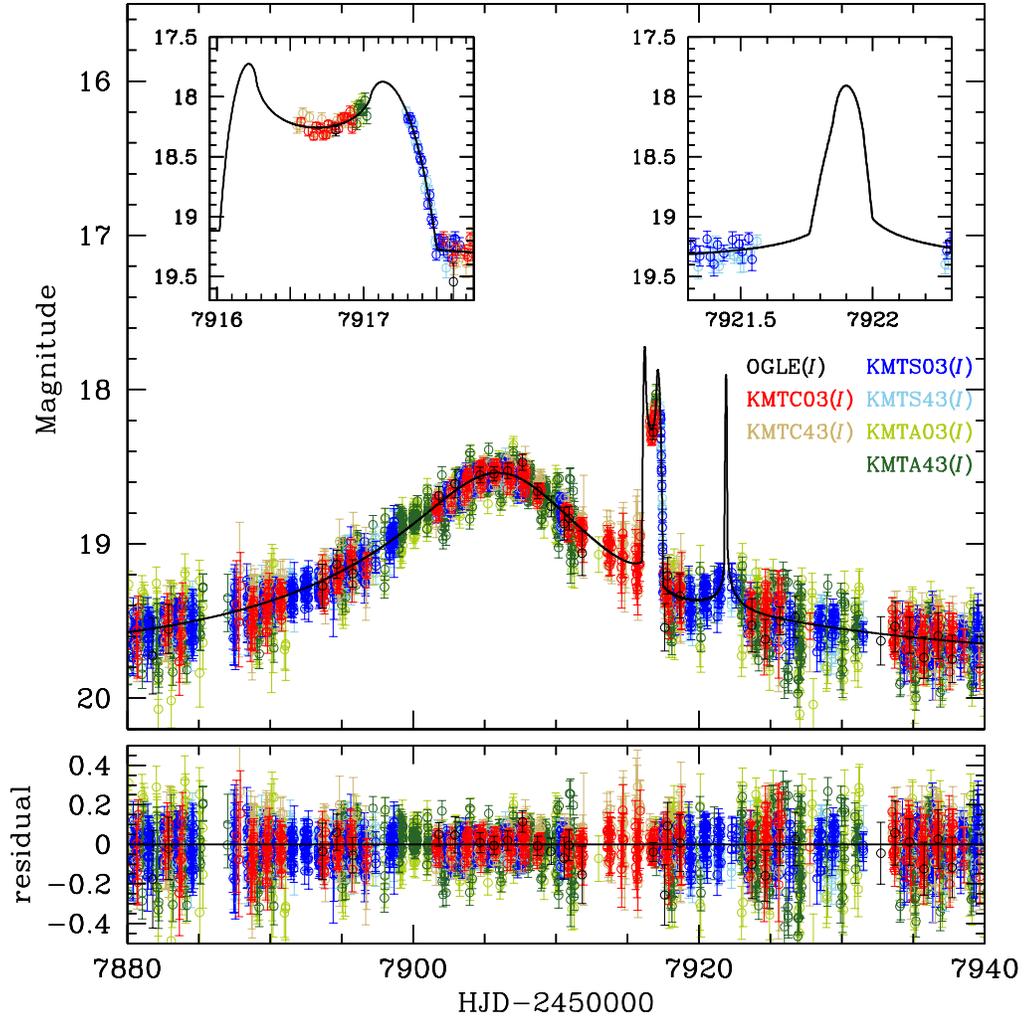}
    \caption{Best-fit light curve of OGLE-2017-BLG-1049 without MOA data. The best-fit light curve is illustrated as the solid black line and the data sets of each observatory are marked in the individual colors. The best-fit parameters for this case are given in the first column of Table 2.}
    \label{fig:fig1}
\end{figure}

\begin{figure}
	\plotone{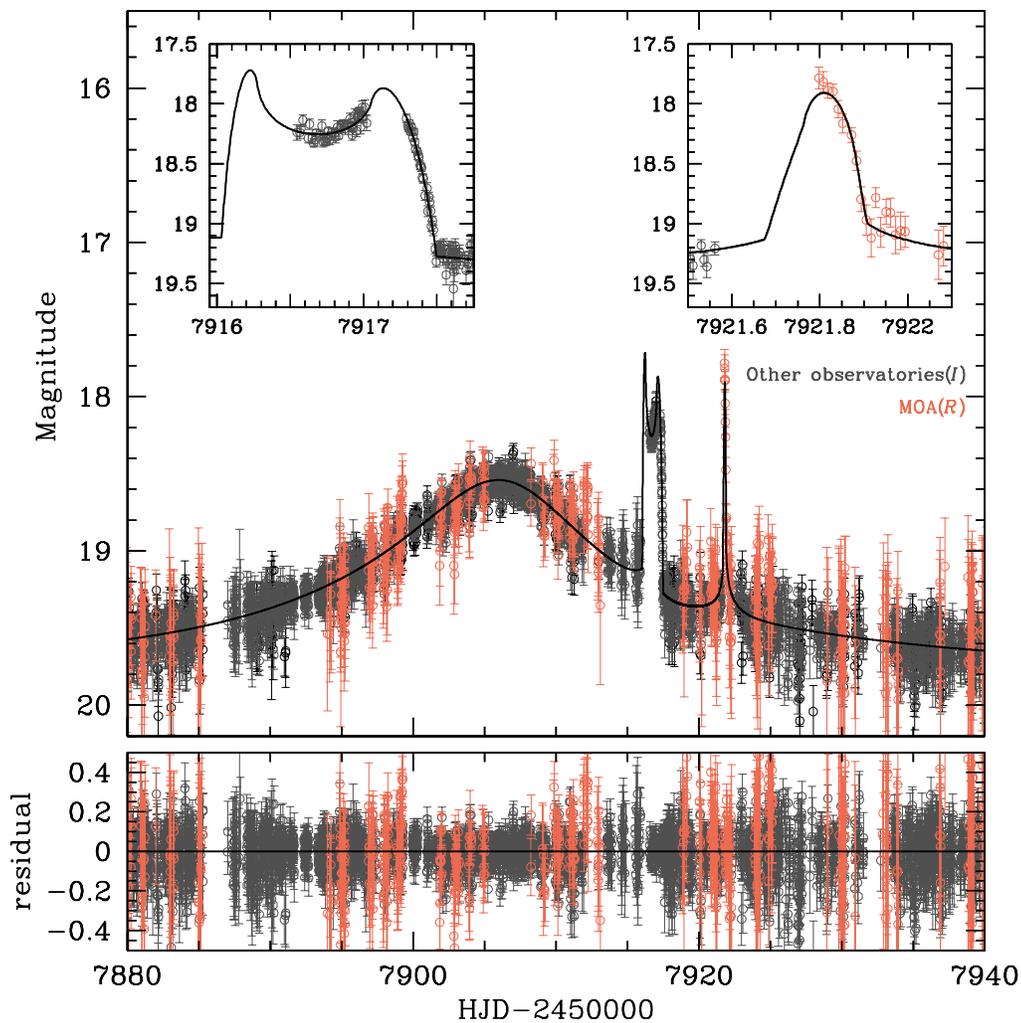}
    \caption{Best-fit light curve of OGLE-2017-BLG-1049 with MOA data. The best-fit light curve is illustrated as the solid black line. The MOA data set expressed in light red and other observatories are so as in grey. The best-fit parameters for this case are given in the second column of Table 2.}
    \label{fig:fig2}
\end{figure}

The KMTNet data show a clear short-duration caustic crossing feature around HJD$'\sim 7917$, as shown in Figure ~\ref{fig:fig1}. This implies that the lens is likely a binary system having a low-mass companion (Gould\citealt{gould2000}, Gould\citealt{gould2001}, Gaudi\citealt{gaudi2012}). Thus, we conduct standard binary lens modeling, which requires seven parameters including three single lens parameters $(t_{0},u_{0},t_{\rm E})$, three binary lens parameters $(s,q,\alpha)$, and $\rho$. Here $t_{0}$ is the time of the closest source approach to the lens, $u_{0}$ is the separation between the source and the lens at $t_{0}$ which is normalized by the angular Einstein ring radius ($\theta_{\rm E}$), which corresponds to the total lens mass, $t_{\rm E}$ is the time duration of crossing $\theta_{\rm E}$, $s$ is the projected separation between the lens components in units of $\theta_{\rm E}$, $q$ is the mass ratio of the lens components, $q=m_{2}/m_{1}$, in which $m_{1}$ and $m_{2}$ are the masses of the host and its companion, respectively, $\alpha$ is the angle between the source trajectory and the binary axis, and $\rho=\theta_{\star} / \theta_{\rm E}$ is the normalized source radius, where $\theta_{\star}$ is the angular size of the source star. In addition to the geometric parameters, there are two flux parameters for each observatory, the source flux $f_{\rm s}$ and the blended flux $f_{\rm b}$. At a given time $t$, the observed fluxes $F(\rm t)$ is 

\begin{equation}
F(t)={f}_{\rm s}A(t)+{f}_{\rm b},
\end{equation}
where $A(t)$ is the magnification as a function of time.

We carry out a grid search on three parameters ($\log s, \log q,  \alpha$) for local $\chi^{2}$ minima using a Markov Chain Monte Carlo (MCMC) algorithm. We investigate $-1 \leq \log s \leq 1$, $-4.0 \leq \log q \leq 0$ and, $0 \leq \alpha \leq 2\pi$, with $100$, $100$, and $21$ uniform grid steps, respectively. In the grid search, $\log(s,q)$ are fixed, while the remaining parameters are allowed to vary in the chain. 
From the local grid search, we found four possible local $\chi^{2}$ minima at $(s,q)= (1.32,9.0 \times10^{-3}), (1.20,1.0 \times10^{-3}), (0.81,1.5 \times10^{-2})$ and, $(0.78,4.4 \times10^{-2})$. We then seed these four sets of local solutions into the MCMC for which all parameters are allowed to vary. During the MCMC, the two close solutions converge to $(s,q)=(0.80,2.22\times10^{-2})$, while the two wide solutions converge to $(s,q)= (1.32,9.53\times10^{-3})$. However, the $\chi^{2}$ of the final close solution is much larger than one of the final wide solution by $\sim1570$. Finally, this event has only one wide solution.

\begin{figure}
\plotone{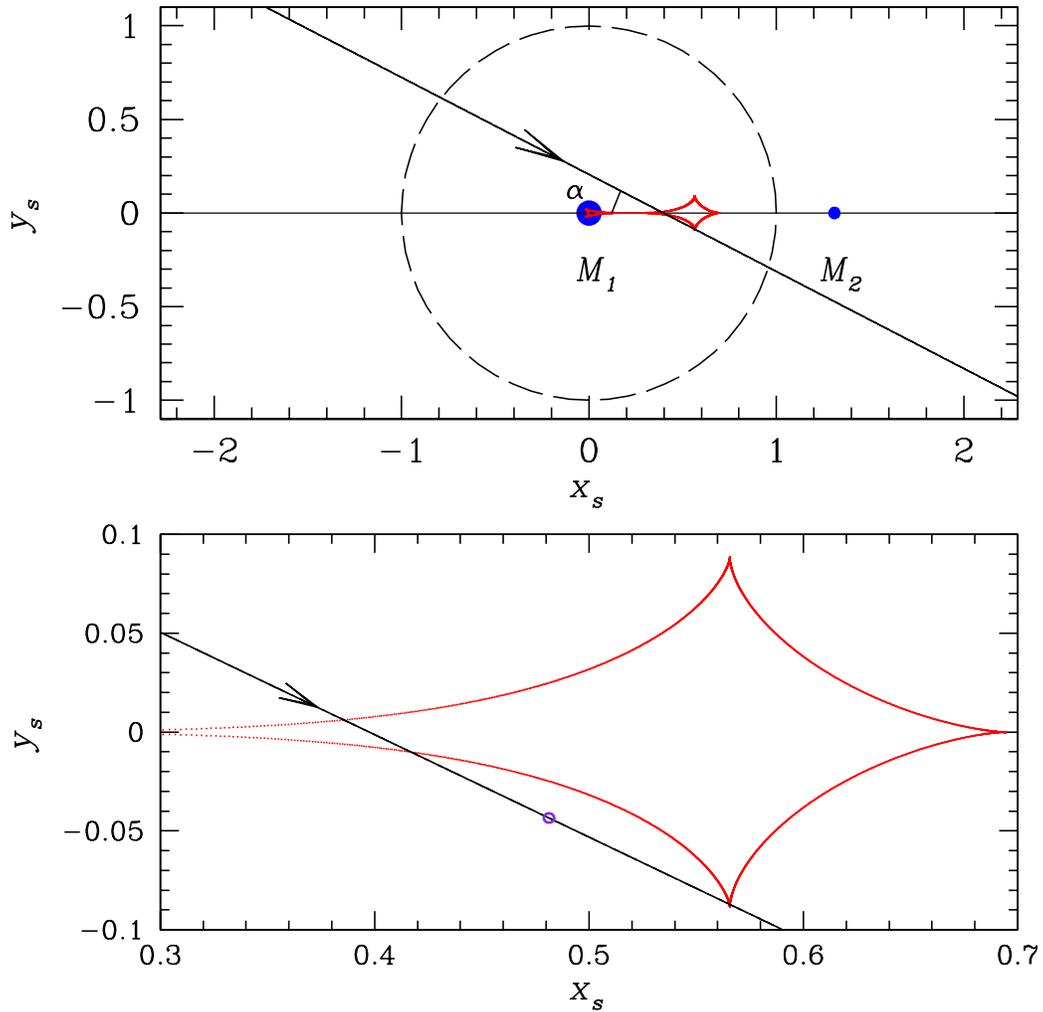}
\caption{Geometry of the best-fit model with MOA data. The lensing parameters for the geometry can be found in the second column of Table 2. The coordinates $(x_{\rm s},y_{\rm s})$ are normalized by the angular Einstein radius corresponding to the total mass of the lens. The black dashed and solid lines in the upper panel indicate the angular Einstein ring and the source star trajectory, respectively, and $\alpha$ is the angle between the source trajectory and the binary axis. The caustics are plotted in red. $M_{1}$ and $M_{2}$ are the host and planetary companion, respectively. The lower panel shows the zoomed image of the caustic and cusp crossing regions. The purple open circle indicates the normalized source size.}
\label{fig:fig3}
\end{figure}

Figure ~\ref{fig:fig2} shows the best-fit light curve of the planetary event OGLE-2017-BLG-1049. The best-fit lensing parameters are presented in Table 2. As can be seen from Figures 1 and 2, the two major perturbations to the light curve were caused by the resonant caustic (Erdl \& Schneider\citealt{erdl1993},Gaudi\citealt{gaudi2012}), which is induced by $(s,q)=(1.32,\ 9.53 \times10^{-3})$ and its geometry is shown in Figure ~\ref{fig:fig3}. We initially perform the MCMC modeling with OGLE and KMTNet data only. As a result, the first perturbation HJD$'\sim7917$ was caused by a caustic crossing and was well covered by KMT data, whereas the second perturbation around HJD$'\sim7922$ was caused by a cusp crossing, but unfortunately, due to this gap in the data, the second perturbation looks \textquotedblleft doubtful\textquotedblright \ (see Figure ~\ref{fig:fig1}), even though the first perturbation was well fitted.

However, after reviewing the MOA observing sequence, it was determined that MOA had observed this event during the data gap, although they did not alert this event, as mentioned in Section 2. Thus, we again carried out the MCMC modeling including MOA data set. As a result, we find that the two modeling results with and without the MOA data are nearly the same at the $1\sigma$ level, which means the $68\%$ uncertainty range from the best-fit model (see Table 2). The best-fit light curve for modeling with the MOA data confirms that the second perturbation is real, although they have some correlated residuals at the beginning of the night.

\begin{figure}
	\plotone{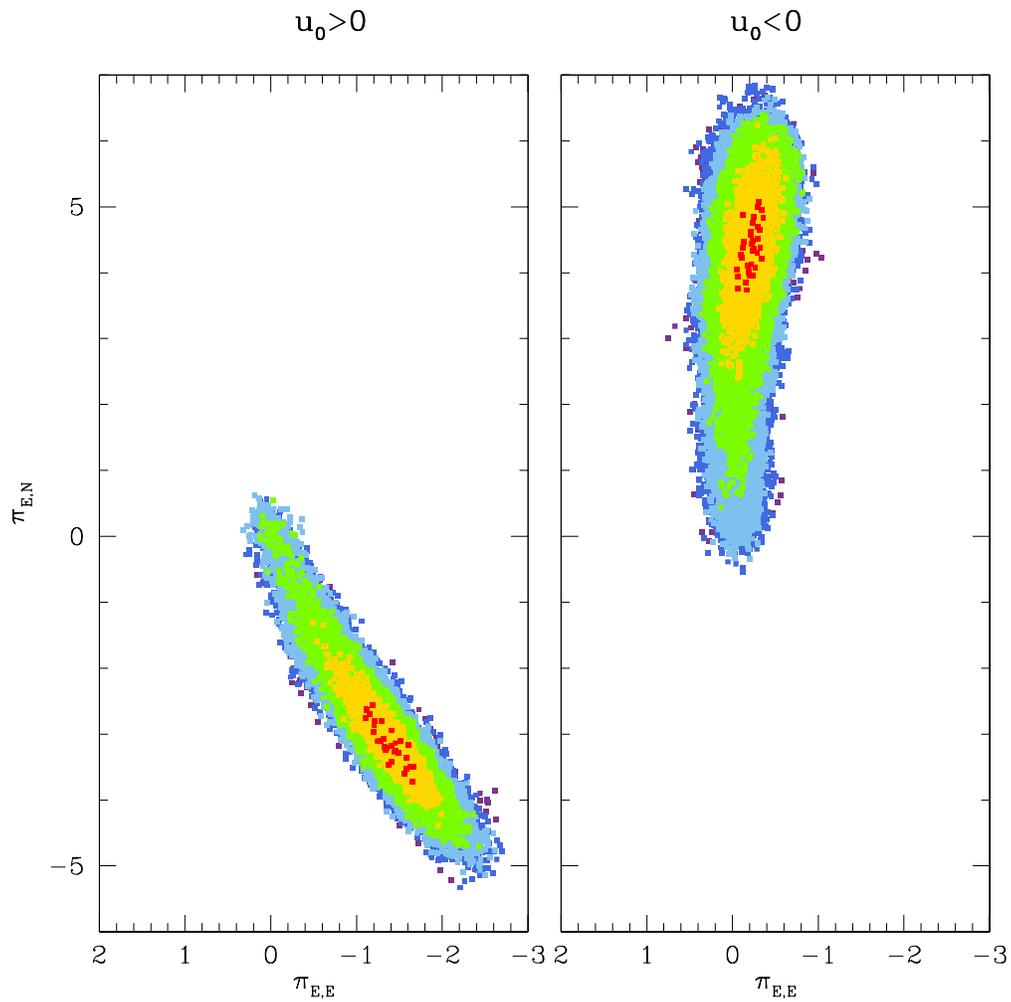}
    \caption{$\Delta\chi^{2}$ distribution of $(\pi_{\rm E,E}, \pi_{\rm E,N})$ for two best-fit parallax+orbital models, $u_{0}>0$ and $u_{0}<0$. The colors of red, yellow, green, light blue, blue, and purple represent regions with $\Delta\chi^{2} < 1, 4, 9, 16, 25$ and $36$ from the best-fit value, respectively.}
    \label{fig:fig4}
\end{figure}

We also conduct the binary lens modeling with high-order parameters including the microlens parallax (Gould\citealt{gould1992},\citealt{gould1994b}) and orbital motion (Dominik\citealt{dominik1998}). In microlensing, the parallax effect is caused by the Earth's orbital motion during the course of the event. Thus, it is a vector which is composed of the parallel and perpendicular properties to the Earth's projected acceleration at the peak of the event which are defined as $\pi_{\rm{E}}=(\pi_{\rm{E,N}}, \pi_{\rm{E,E}})$ (Gould et al.\citealt{gould1994b}, Udalski et al.\citealt{udalski2018}). The orbital motion effect is caused by the orbital motion of the binary lens under the assumption of the linear orbital motion. It is described by two parameters $(ds/dt, d\alpha/dt)$, which are the change rates of the binary separation and the orientation angle of the binary axis, respectively. However, it is found that the orbital motion effect can mimic the parallax effect (Batista et al.\citealt{batista2011}, Skowron et al.\citealt{skowron2011}, Han et al.\citealt{han2016}). Hence, we model the event with both two effects. From this, we find that the best-fit parallax+orbital model, i.e. the case of $u_{0}>0$, improves the fit by only $\Delta\chi^{2} \lesssim 5$ compared to the standard model. This means that the high-order effects were not detected. Figure ~\ref{fig:fig4} shows that the magnitude of the microlens parallax is not well constrained, but the direction is. Therefore, we perform a Bayesian analysis to determine the physical parameters of the lens system, which will be described in Section 4.

\section{physical parameters}
\subsection{Angular Einstein radius}
Thanks to the detection of the caustic crossing feature, we were able to measure the normalized source radius, $\rho=\theta_{\star}/\theta_{\rm E}$, from the light curve modeling. In order to determine the angular Einstein radius $\theta_{\rm E}$, we must measure the source radius $\theta_{\star}$, which is determined from the intrinsic brightness and color of the source star. As described in Yoo et al.\citet{yoo2004}, the intrinsic brightness and color of the source can be determined from the offset between the red giant clump centroid (RGC) and the source in the instrumental color-magnitude diagram (CMD) under the assumption that they experience the same amount of extinction, resulting in

\begin{eqnarray}
(V-I, I)_{\rm s,0}=(V-I, I)_{\rm s}-(V-I, I)_{\rm RGC}\nonumber\\
+(V-I, I)_{\rm RGC,0}.
\label{eqn:vii}
\end{eqnarray}

\begin{figure}
	\plotone{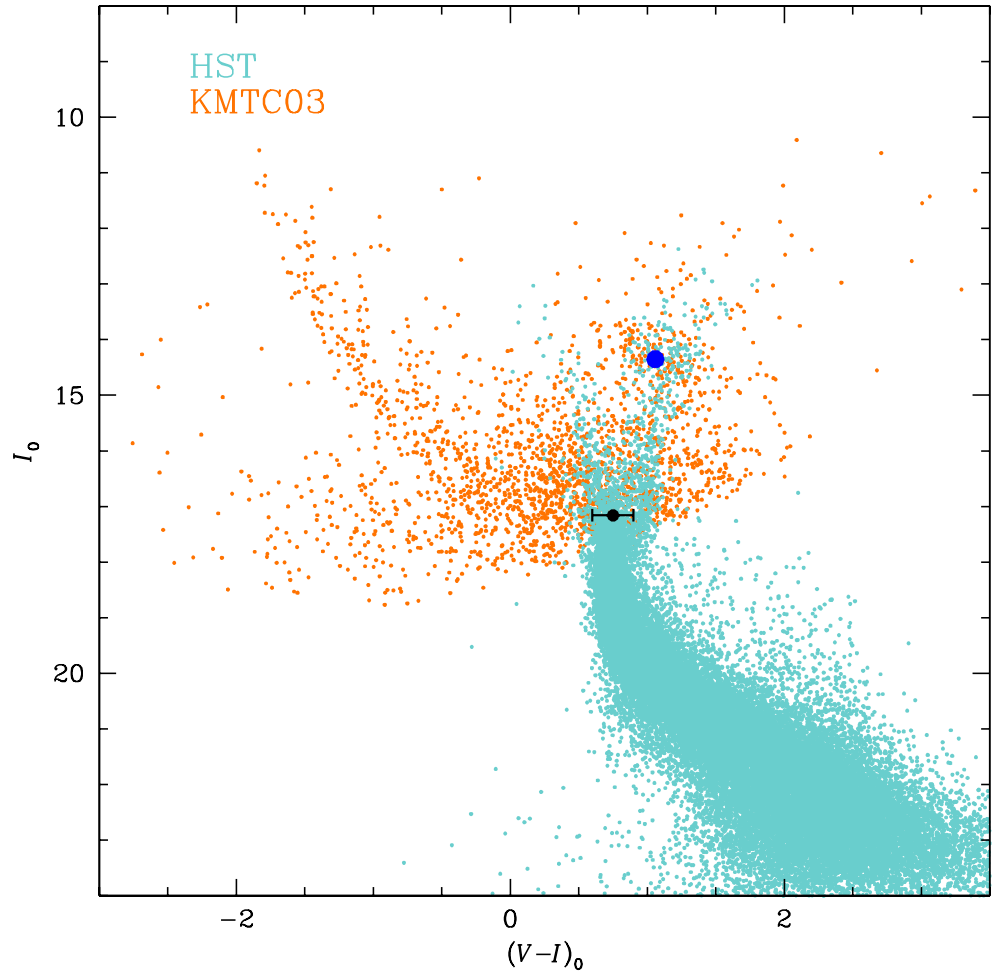}
    \caption{Color-magnitude diagram (CMD) of field stars in a 120$''$ square centered on the event OGLE-2017-BLG-1049. KMTNet CMD is shifted by the HST CMD. The blue and black dot indicate the positions of the Red Giant Clump centroid (RGC) and the source.}
    \label{fig:fig5}
\end{figure}

However, the quality of KMTNet V-band data for this event was too poor to estimate a reliable color value; hence, we estimate the intrinsic V-I color $(V-I)_{0}$ using the Hubble Space Telescope CMD from Holtzman et al.\citet{holtzman1998} following Section 5.2 of Bennett et al.\citet{bennett2008}. The KMTNet CMD aligned to the HST CMD is shown in Figure ~\ref{fig:fig5}. We adopt $(V-I,I)_{\rm RGC,0}=(1.06, 14.34)$ from Nataf et al\citet{nataf2013}. As a result, we find $(V-I, I)_{\rm s,0} = (0.75 \pm 0.15, 17.14 \pm 0.01)$ and by using $VIK$ color-color relation from Bessell \& Brett\citet{bessell1998} and the color-surface brightness relation from Kervella et al.\citet{kervella2004}, we find that $\theta_{\star}=1.25\pm 0.20 \ \mu{\rm as}$. Note that Kervella et al.\citet{kervella2004} only applies to dwarf stars and subgiants. Judging by the color and source size, it indicates that the source is likely a late G-type sub-giant star. The determined $\rho$ and $\theta_{\star}$ yield the angular Einstein radius of the lens system,

\begin{equation}
\theta_{\rm E}=\frac{\theta_{\star}}{\rho}=0.52\pm0.11\, {\rm mas}.
\end{equation}

Since the source is located in the region of the CMD where the giant branch separates from the main sequence, we repeat our calculations with a redder $(V-I)_{\rm s,0}=1.0$ to check the impact of the color on the Bayesian fit. The difference in $\theta_{\star}$ is less than $3\%$; therefore, we adopted $(V-I)_{\rm s,0}=0.75$ which is the more likely value. The parameters related to the source star are summarized in Table 3.

\subsection{Lens properties}
\begin{figure}
	\plotone{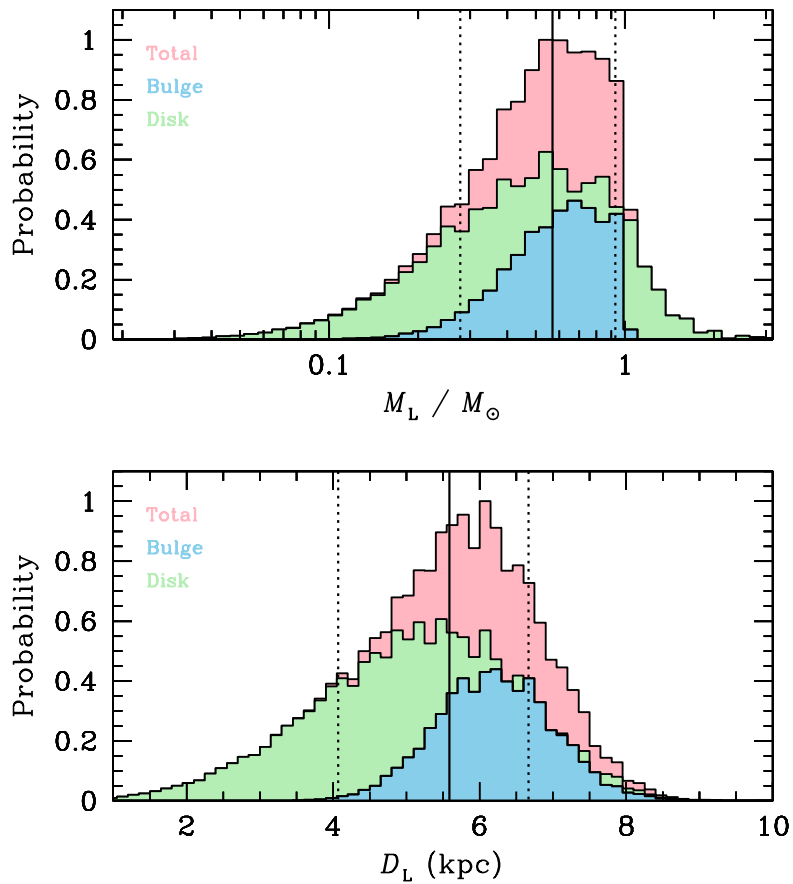}
    \caption{Probability distributions of the lens mass (host star) and distance to the lens derived from the Bayesian analysis. The vertical black solid line represents the median value and the two vertical dotted lines represent the confidence intervals of 68$\%$.}
    \label{fig:fig6}
\end{figure}

In order to determine the mass and distance of the lenses, measurement of the angular Einstein radius $\theta_{\rm E}$ and the microlens parallax $\pi_{\rm E}$ are required. This is because

\begin{equation}
M_{\rm L}=\frac{\theta_{\rm E}}{\kappa\pi_{\rm E}}, \ \ D_{\rm L}=\frac{\rm au}{\pi_{\rm E}\theta_{\rm E}+\pi_{\rm S}}, 
\label{lens1}
\end{equation}
where $\kappa \equiv 4G / c^{2} \backsimeq 8.144\, {\rm mas}\, M_{\odot}^{-1}$, $\pi_{\rm S}={\rm au}/D_{\rm S}$ is the source parallax (Gould\citealt{gould2000}). Here we adopt $D_{\rm S}=7.70\ {\rm kpc}$ from Nataf et al.\citet{nataf2013}. However, as mentioned above, there is no reliable parallax detection for OGLE-2017-BLG-1049. Hence, we perform a Bayesian analysis to determine $M_{\rm L}$ and $D_{\rm L}$. The Bayesian analysis is conduced by the same procedure as described in Jung et al.\citet{jung2018} except for a Galactic model, which is based on three priors of the velocity distribution, mass function, and matter density profile of the Galaxy. Here we use a new Galactic model based on more recent data and scientific understanding. The new Galactic model includes the disk density profile from Robin-based Bennett model, disk velocity dispersion (Bennett et al.\citealt{bennett2014}), and the bulge mean velocity and dispersions from $\it{Gaia}$, while the bulge density profile is the same one as the previous model (i.e., Jung et al.\citealt{jung2018}). In the previous Galactic model, the bulge mean velocity was zero. With these three priors, we generate a several tens of millions artificial microlensing events and investigate the probability distributions of $M_{\rm L}$ and $D_{\rm L}$ using measured $t_{\rm E}$ and $\theta_{\rm E}$. From this, we find that the mass and distance of the host star are
\begin{equation}
M_{\rm h}=0.55^{+0.36}_{-0.29}\ M_{\odot}, \ \ \ D_{\rm L} = 5.67^{+1.11}_{-1.52}\ {\rm kpc},
\end{equation}
with the values being the median values of the probability distributions and the errors indicating the $68\%$ confidence intervals (see Figure ~\ref{fig:fig6}).
We then estimate the mass $M_{\rm p}$ and the projected separation $a_{\perp}$ of the planetary companion, which are determined by $M_{\rm p}=qM_{\rm h}$ and $a_{\perp}=sD_{\rm L}\theta_{\rm E}$:
\begin{equation}
M_{\rm p}=5.53^{+3.62}_{-2.87}\, M_{\rm Jup}, \ \ \ a_{\perp}=3.92^{+1.10}_{-1.32}\ \rm au.
\end{equation}
Accordingly, the lens system of OGLE-2017-BLG-1049 is likely to be an M dwarf star hosting a super Jupiter-mass planet. Considering that the snow line of a host star scales as $a_{\rm snow}=2.7 (M/M_{\odot})\, {\rm{au}}$ (Kennedy \& Kenyon\citealt{kennedy2008}), the snow line of the lens system is $a_{\rm{snow}}\simeq 1.49\ \rm{au}$. This indicates that the super Jupiter-mass planet lies beyond the snow line. All physical parameters of the lens found from the Bayesian analysis are listed in Table 4.

We additionally conduct a Bayesian analysis including additional parallax constraints. The nominal, best-fit value of the parallax is $\pi_{\rm{E}} \sim 4$, which implies $M_L \sim 0.015 M_{\odot}$ and $D_L \sim 500\,$ pc. Such a result would be quite remarkable. However, the constraints on $\pi_{\rm{E}}$  are very weak (see Figure ~\ref{fig:fig4}). Therefore, the Bayesian priors dominate over the parallax, and including the parallax hardly changes the results (Table 4).

Using the measured $\theta_{\rm{E}}$ and $t_{\rm E}$, we estimate the relative lens-source proper motion to be $\mu_{\rm{rel}}=\theta_{\rm{E}}/t_{\rm{E}} = 6.66 \pm 1.35\, {\rm{mas}\, yr^{-1}}$, which is a value consistent with disk objects. The Bayesian analysis supports a disk lens with $(D_{\rm{L}}\sim 5.7 \rm{ \ kpc})$.

\section{summary}
We analysed the planetary lensing event OGLE-2017-BLG-1049 with a short-timescale caustic crossing feature detected by KMTNet high-cadence observations. We adopted a best-fit model light curve including MOA data using a Bayesian analysis with the constraints $\rm t_{E}+\theta_{E}$. From this, we found a star with mass $M_{\rm h}=0.55^{+0.36}_{-0.29}\ M_{\odot}$ hosting a giant planet with a mass $M_{\rm p}=5.53^{+3.62}_{-2.87}\ M_{\rm Jup}$. The projected star-planet separation is $a_{\perp}=3.92^{+1.10}_{-1.32}\ {\rm au}$, indicating that the giant planet lies beyond the snow line. Thus, the lens system is likely to be an M dwarf star. However, it is also possible that the host star is a K or even G dwarf.

\begin{figure}
	\plotone{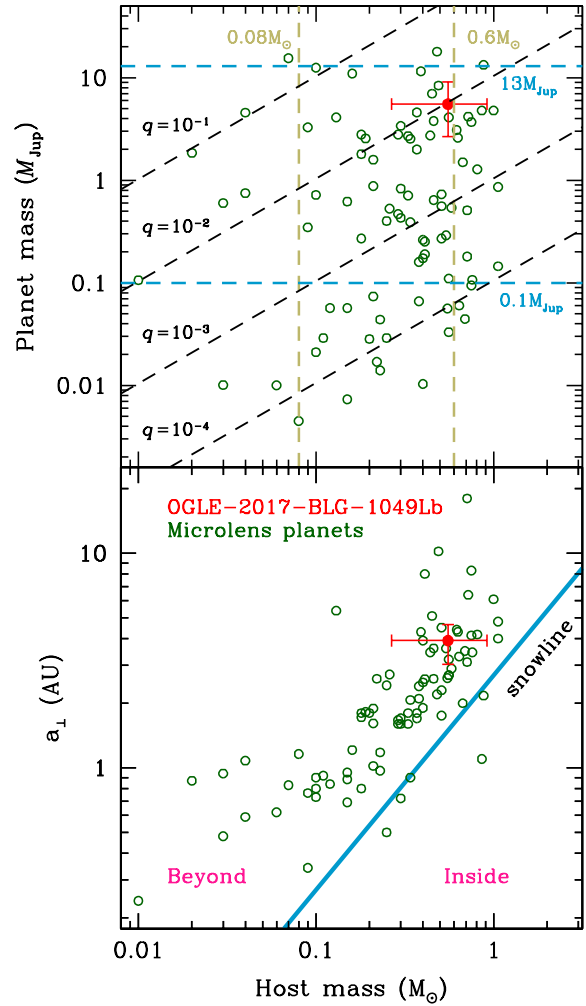}
    \caption{Distribution of microlensing exoplanets from the NASA exoplanet archive\textsuperscript{\ref{nasa}}. The upper panel shows the distribution of the planet mass and the host mass. The yellow vertical dashed lines represent the mass limits of M dwarf stars $(0.08\sim0.6M_{\odot})$ and the blue dashed lines represent the mass limit of giant planets $(0.1\sim13M_{\rm Jup})$. The lower panel shows the distribution of the host star and the projected semi-major axis with the snow line $a_{\rm snow}=2.7 (M/M_{\odot})\, {\rm au}$. The red dot indicates the planet discovered in this paper, which is shown with uncertainties, but all other planets (green dots) are not.}
    \label{fig:fig7}
\end{figure}

Figure ~\ref{fig:fig7} shows OGLE-2017-BLG-1049Lb compared to other known microlensing exoplanets. The uncertainties are the 1-sigma uncertainties derived from Bayesian analysis and are typical of such Bayesian mass estimates. The distributions suggest that the distribution of giant planets is uniform from M to G dwarfs, with no noticeable decline in the frequency of giant planets around M dwarfs. However, host mass measurements are necessary in order to disentangle the underlying population from the Bayesian priors.

In the case of OGLE-2017-BLG-1049, it should be possible to measure the flux from the host star in a few years from now. The source-lens relative proper motion is about $\mu_{\rm rel}\sim7 \rm{mas \ yr^{-1}}$, implying that, by 2026, the two stars will be separated by $60$ mas. Thus, the host star can probably be resolved by existing 10 meter class telescopes at that time, or (if necessary) by the extremely large telescopes currently under construction.

\section*{acknowledgements}

Work by Y. H. Kim was supported by the KASI (Korea Astronomy and Space Science Institute) grant 2020-1-830-08. This research has made use of the KMTNet system operated by the Korea Astronomy and Space Science Institute (KASI) and the data were obtained at three host sites of CTIO in Chile, SAAO in South Africa, and SSO in Australia. The OGLE project has received funding from the National Science Centre, Poland, grant MAESTRO 2014/14/A/ST9/00121 to A.U. The MOA project was supported by JSPS KAKENHI grant No. JSPS24253004, JSPS26247023, JSPS23340064, JSPS15H00781, JP17H02871, and JP16H06287. This work has made use of the NASA Exoplanet Archive, which is operated by the California Institute of Technology, under contract with the National Aeronautics and Space Administration under the Exoplanet Exploration Program. Work by C.H. was supported by the grants of National Research Foundation of Korea (2017R1A4A1015178 and 2019R1A2C2085965).


\begin{deluxetable}{lcccccc}
\tablecolumns{4} \tablewidth{0pc} \tablecaption{\textsc{Data and error rescaling parameters}}
\tablehead{\colhead{Observatory (Band)} & \colhead{$N_{\rm data}$} & \colhead{{\it k}} & \colhead{$\sigma_{0}\ (\rm mag)$}}
\startdata
OGLE       ({\it I}) & 235  & 1.450 & 0.002 \\
KMTC BLG03 ({\rm I}) & 1238 & 1.189 & 0.000 \\
KMTC BLG43 ({\rm I}) & 719  & 1.354 & 0.000 \\
KMTS BLG03 ({\rm I}) & 1746 & 1.227 & 0.000 \\
KMTS BLG43 ({\rm I}) & 1677 & 1.247 & 0.000 \\
KMTA BLG03 ({\rm I}) & 1404 & 1.021 & 0.000 \\
KMTA BLG43 ({\rm I}) & 1418 & 1.038 & 0.000 \\
MOA        ({\it R}) & 3466 & 0.961 & 0.000 \\
\enddata
\label{table:table1}
\tablecomments{$N_{\rm data}$ is the number of data points. The definition of parameters can be seen from Eq.~\ref{eqn:yee}.}
\end{deluxetable}

\begin{deluxetable}{lcccc}
\tablecolumns{3} \tablewidth{0pc} \tablecaption{\textsc{Best-fit parameters and their $68\%$ uncertainty range from MCMC}}
\tablehead{\colhead{Parameters}           &      \colhead{w/o MOA}            &         \colhead{\textbf{w/ MOA}}} \startdata
$\chi^2/\rm{dof}$            & 8335.887/8416           &   \textbf{11796.306/11880}    \\
$t_0$ $(\rm{HJD}^{\prime})$  & 7906.413 $\pm$ 0.029    &   \textbf{7906.453 $\pm$ 0.027} \\
$u_{0}$                      & 0.180    $\pm$ 0.002    &   \textbf{0.183    $\pm$ 0.002} \\
$t_{\rm E} (\rm{days})$      & 28.861   $\pm$ 0.283    &   \textbf{28.652   $\pm$ 0.239} \\
$s$                          & 1.320    $\pm$ 0.003    &   \textbf{1.320    $\pm$ 0.003} \\
$q$ $(10^{-3})$              & 9.645    $\pm$ 0.471    &   \textbf{9.534    $\pm$ 0.386} \\
$\alpha (\rm{rad})$          & 0.474    $\pm$ 0.004    &   \textbf{0.478    $\pm$ 0.003} \\
$\rho (10^{-3})$             & 2.384    $\pm$ 0.052    &   \textbf{2.385    $\pm$ 0.049} \\
$f_{\rm s,OGLE}$             & 0.094    $\pm$ 0.001    &   \textbf{0.095    $\pm$ 0.001} \\
$f_{\rm b,OGLE}$             & 0.103    $\pm$ 0.001    &   \textbf{0.180    $\pm$ 0.002} \\
\enddata
\tablecomments{HJD$\prime$ = HJD - 2450000 days. The values in bold represent the adopted best-fit model.}
\label{table:table2}
\end{deluxetable}

\begin{deluxetable}{lrrrr}
\tablecolumns{2} \tablewidth{0pc} \tablecaption{\textsc{Source parameters}}
\tablehead{\colhead{Parameters} & & & & \colhead{Values}}
\startdata
$(V-I)_{\rm s,0}$               & & & & 0.75   $\pm$ 0.15  \\
$I_{\rm s,0}$                   & & & & 17.14  $\pm$ 0.01  \\
$\theta_{\star} \ (\mu \rm as)$  & & & & 1.25  $\pm$ 0.20 \\
\enddata
\label{table:table3}
\end{deluxetable}

\begin{deluxetable}{lcccccc}
\tablecaption{\textsc{Lens parameters}}
\tablehead{\multicolumn{1}{c}{Parameter} &  \multicolumn{1}{c}{\boldmath{$t_{\rm{E}} + \theta_{\rm{E}}$}}  &\multicolumn{2}{c}{$t_{\rm{E}}+\theta_{\rm{E}}+\pi_{\rm{E}}$} \\  
& & $u_{0}>0$ & $u_{0}<0$} \startdata
$\theta_{\rm{E}} (\rm{mas})$   & \textbf{0.52 $\pm$ 0.11}           & 0.48 $\pm$ 0.10        & 0.49 $\pm$ 0.10 \\
$\mu_{\rm{rel}} (\rm{mas/yr})$ & \textbf{6.66 $\pm$ 1.35}           & 5.97 $\pm$ 1.21        & 6.01 $\pm$ 1.21 \\
$M_{\rm{h}} (M_{\odot})$       & \boldmath{$0.55^{+0.36}_{-0.29}$}  & $0.59^{+0.33}_{-0.26}$ & $0.48^{+0.39}_{-0.27}$ \\
$D_{\rm{L}} \ (\rm{kpc})$      & \boldmath{$5.67^{+1.11}_{-1.52}$}  & $6.09^{+0.95}_{-1.14}$ & $5.54^{+1.25}_{-1.72}$ \\
$a_{\perp} \ (\rm{au})$        & \boldmath{$3.92^{+1.10}_{-1.32}$}  & $4.00^{+1.02}_{-1.10}$ & $3.64^{+1.59}_{-1.06}$ \\
\enddata
\tablecomments{$t_{\rm{E}}$, $\theta_{\rm{E}}$, and $\pi_{\rm{E}}$ are the constraints used in Bayesian analysis. The values in bold represent the preferred best-fit odel.}
\label{table:table4}
\end{deluxetable}

\end{document}